\documentclass{caosp}

\usepackage{graphicx}

\articleNo{123}
\pubyear{2005}
\volume{35}
\volnumber{3}
\firstpage{1}
\received{May 1, 2005}
\accepted{August 28, 2005}

\begin{document}

\htitle{A discovery of low-contrast fields in slow rotators?}
\hauthor{O. Kochukhov}

\title{Zeeman split lines in CP stars: 
a discovery of low-contrast fields in slow rotators?}


\author{O. Kochukhov}

\institute{
Department of Astronomy and Space Physics, 
Uppsala University, Box 515, SE-751 20 Uppsala, Sweden
}

\date{December 1, 2007}

\maketitle

\begin{abstract}
We show that high-resolution observations of resolved Zeeman split 
lines can be used to obtain new constraints on the stellar magnetic field 
geometry. In
particular, the contrast of the field strength distribution over the stellar
surface can be deduced from the differential measurement of the second moment
of the $\pi$ and $\sigma$ Zeeman components. Our analysis of the triplet lines in
slowly rotating cool
magnetic CP stars uncovers a surprisingly homogeneous field structure,
inconsistent with any low-order multipolar geometry.
\end{abstract}

\section{Introduction}

A number of slowly rotating magnetic A and B stars have surface magnetic fields strong
enough to produce observable Zeeman splitting of line profiles in high-resolution
spectroscopic observations. This effect is routinely used to diagnose the
surface magnetic field strength (e.g. Mathys et al. \cite{mathys}). But, at the same time,
information provided by the shapes of the resolved $\pi$ and $\sigma$ components is
largely neglected. Here we propose
a novel technique to measure the {\it relative width} of the resolved Zeeman components
and to infer an important global field characteristic from the resulting new magnetic observable.

\section{Mean field scatter}

At each point on the magnetic star surface the local line profiles are splitted according
to the magnetic field strength. The $\pi$ component of a triplet line coincides with the
laboratory wavelength, whereas the shift of $\sigma$ components is proportional 
to the 
field modulus and therefore varies over the stellar surface together with the field
strength. Consequently, the
$\sigma$ components are broader than $\pi$ components. Thus, the width of the $\sigma$
components relative to that of $\pi$ components provides a measure of {\it the
contrast of the field strength distribution} over the stellar disk.

Neglecting stellar rotation and using assumptions similar to the derivation of common magnetic 
observables (e.g., Mathys \cite{mathys95}), we can quantify the difference in width of 
the
resolved Zeeman components of a triplet line using measurement of their second moments, 
$M^{(2)}_{\pi,\sigma}$, computed with
respect to the center-of-gravity of each component:
\begin{equation}
M^{(2)}_{\sigma}-M^{(2)}_{\pi}=\left(\frac{\displaystyle e\lambda^2_0}{\displaystyle 4\pi m_e c^2}\right)^2
\bar{g}^2 \langle \delta B^2 \rangle,
\end{equation}
where $\bar{g}$ is the mean Land\'e factor and other variables have their usual meaning.
$\langle \delta B^2 \rangle$ is a new magnetic
observable, {\it mean field scatter}. 
It is primarily sensitive to the field strength
contrast and weakly sensitive to the field orientation.

\begin{figure}[!t]
\centering
\includegraphics[width=6.2cm]{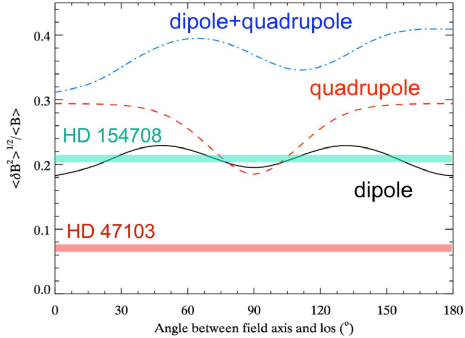}
\caption{Measurements of the mean field scatter in HD\,47103 and HD\,154708 (shaded regions)
are compared with the $\langle \delta B^2 \rangle/\langle B\rangle$ ratio 
expected for multipolar field
geometries viewed from different angle $\alpha$ between the field axis and the line of sight.}
\label{fig1}
\end{figure}

\section{Low-contrast fields in cool Ap stars}

We studied Zeeman split triplet lines in the UVES spectra of cool Ap stars
HD\,47103 and HD\,154708. 
Results of the $\langle \delta B^2 \rangle$ measurements for these stars are summarized in
Fig.~\ref{fig1}. It is evident that large width of $\sigma$ components in HD\,154708 
is consistent with the dipolar field geometry, but the $\sigma$ components are
unexpectedly narrow in HD\,47103. Some other strong-field Ap stars also exhibit
narrow $\sigma$ components. Fig.~\ref{fig1} shows that this observation cannot be
reconciled with {\it any} low-order multipolar field configuration. 

Thus, some slowly rotating cool Ap stars host surprisingly homogeneous
surface magnetic field topologies. This finding might have an important implication for the
theories of formation and evolution of magnetic fields in these stars.

\end{document}